\documentclass[showpacs,amssymb,twocolumn,floatfix,aps,pre,notitlepage]{revtex4-2}

\usepackage{amssymb,amsfonts,amsmath,bm}
\usepackage{graphics,graphicx}
\usepackage{color}
\usepackage{xtab,afterpage}

\usepackage{geometry} 
\begin{document}
\title{Tensor network calculation of boundary and corner magnetization}

\author{Roman \textsc{Krcmar}$^{1}$}
\author{Jozef \textsc{Genzor}$^{1,2}$}
\author{Andrej \textsc{Gendiar}$^{1}$}
\author{Tomotoshi \textsc{Nishino}$^{3}$}

\affiliation{$^1$Institute of Physics, Slovak Academy of Sciences, D\'ubravsk\'a cesta 9, 84511 Bratislava, Slovakia}
\affiliation{$^2$Physics Division, National Center for Theoretical Sciences, National Taiwan University, Taipei 10617, Taiwan}
\affiliation{$^3$Department of Physics, Graduate School of Science, Kobe University, Kobe 657-8501, Japan}

\date{today} 

\begin{abstract}
The Corner Transfer Matrix Renormalization Group (CTMRG) algorithm is modified to measure the magnetization at the boundary of the system, including the corners of the square-shaped lattice. Using automatic differentiation, we calculate the magnetization's first derivative, allowing us to determine the boundary critical exponent $\beta$ accurately.
\end{abstract}

\maketitle

\begin{figure}[htb]
\begin{center}
\includegraphics[width=0.45\textwidth,clip]{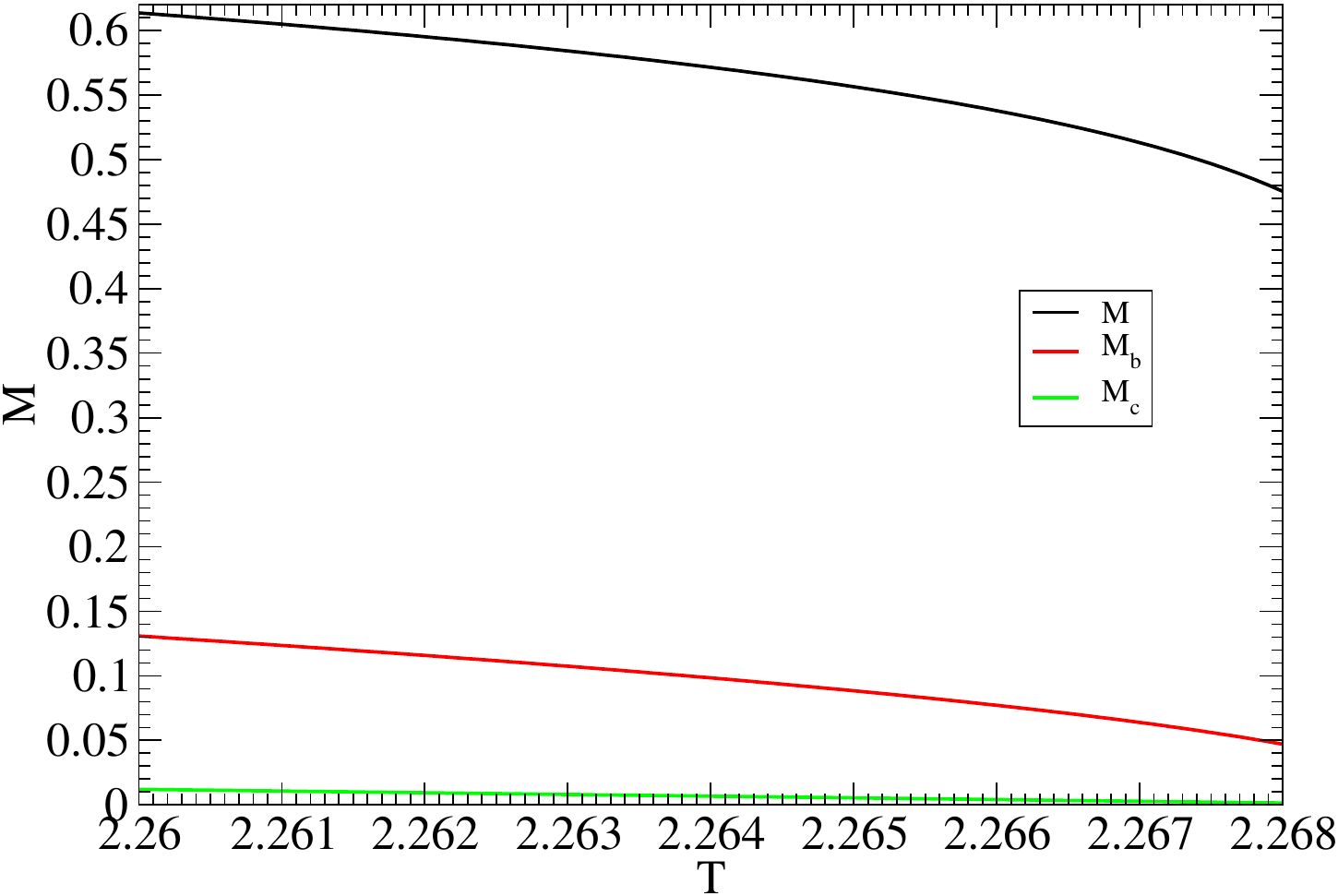}
\caption{Magnetization as a function of the temperature. Bulk magnetization $M$ (black curve), border magnetization $M_b$ (red curve), corner magnetization $M_c$ (green curve). Calculated for $m = 300$.}
\label{fig:mag}
\end{center}
\end{figure}

\section{Introduction}
The study of critical phenomena in two-dimensional systems was revolutionized by the development of Conformal Field Theory (CFT) \cite{BPZ}. The behavior of a system at its boundary, however, can be significantly different from its behavior in the bulk \cite{binder}. This is the domain of Boundary Conformal Field Theory (BCFT), a theory devised by Cardy \cite{cardy84}. Still different behavior is at the corner of the system where boundaries intersect \cite{cardy83}. As first shown by Cardy\cite{cardy83}, critical exponents in a corner can depend on the angle $\theta$ between the adjacent edges. Later studies has shown that boundary critical exponent $\beta_b$ and corner critical exponent $\beta_c$ are connected.~\cite{peschel84,peschel85,peschel88,peschel91,latremoliere,madary}. The relation of these two critical exponents is 
\begin{equation}
\beta_c = \frac\pi\theta\beta_b,
\end{equation}
where $\theta$ is the angle between the edges of the corner.

\begin{figure}[t!]
\begin{center}
\includegraphics[width=0.45\textwidth,clip]{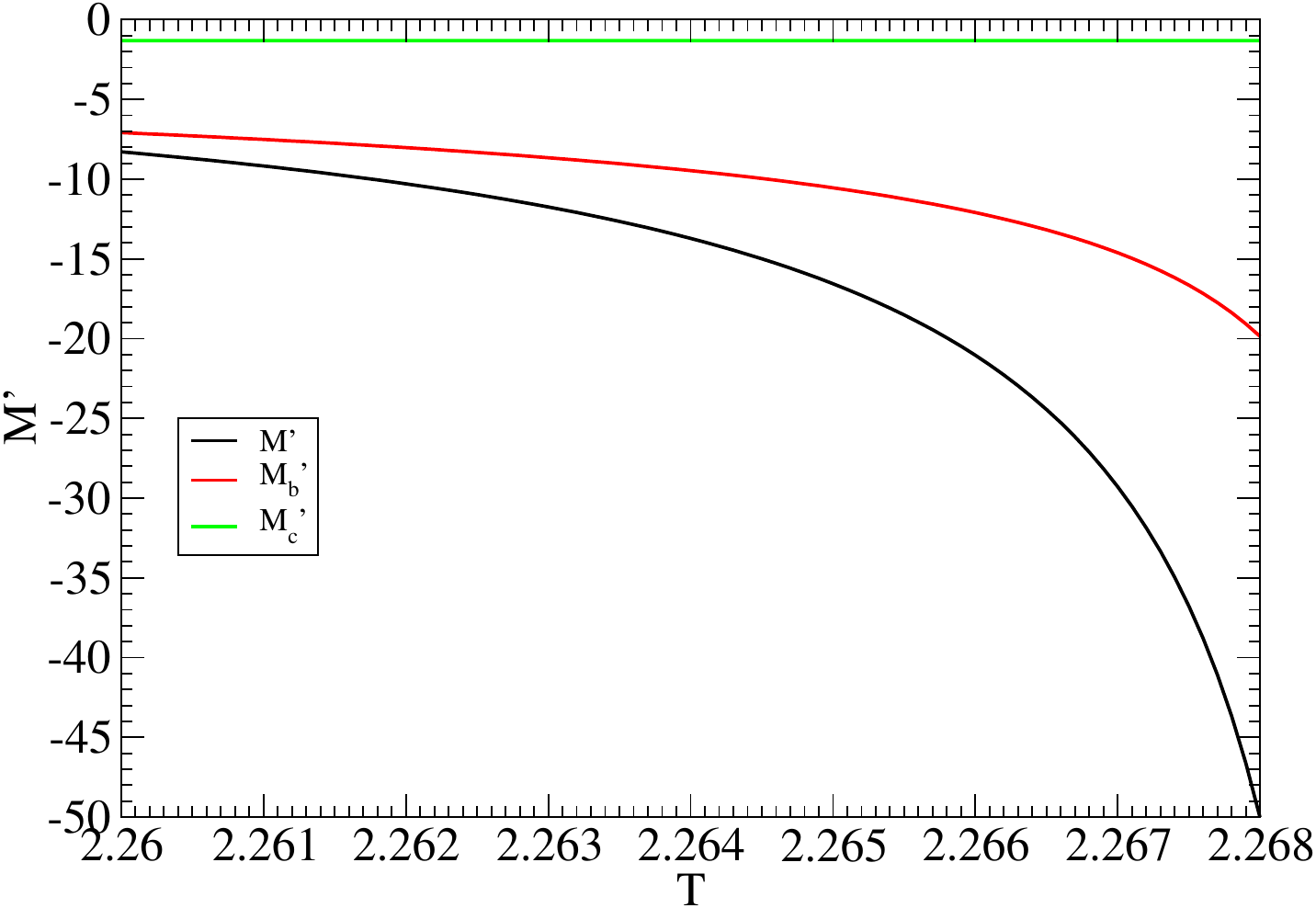}
\caption{First derivative of magnetization as a function of temperature. Bulk magnetization $M^\prime$ (black curve), border magnetization $M_b^\prime$ (red curve), corner magnetization $M_c^\prime$ (green curve). Calculated for $m = 300$.}
\label{fig:der}
\end{center}
\end{figure}
Building on recent progress with the tensor renormalization group (TRG) ~\cite{iino2019,iino2020}, this work extends these powerful numerical techniques to the study of corner critical phenomena. While previous studies focused on bulk or boundary properties, our goal is the direct and precise calculation of the boundary and corner critical exponents, $\beta_b$ and $\beta_c$.

\begin{figure}[t!]
\begin{center}
\includegraphics[width=0.45\textwidth,clip]{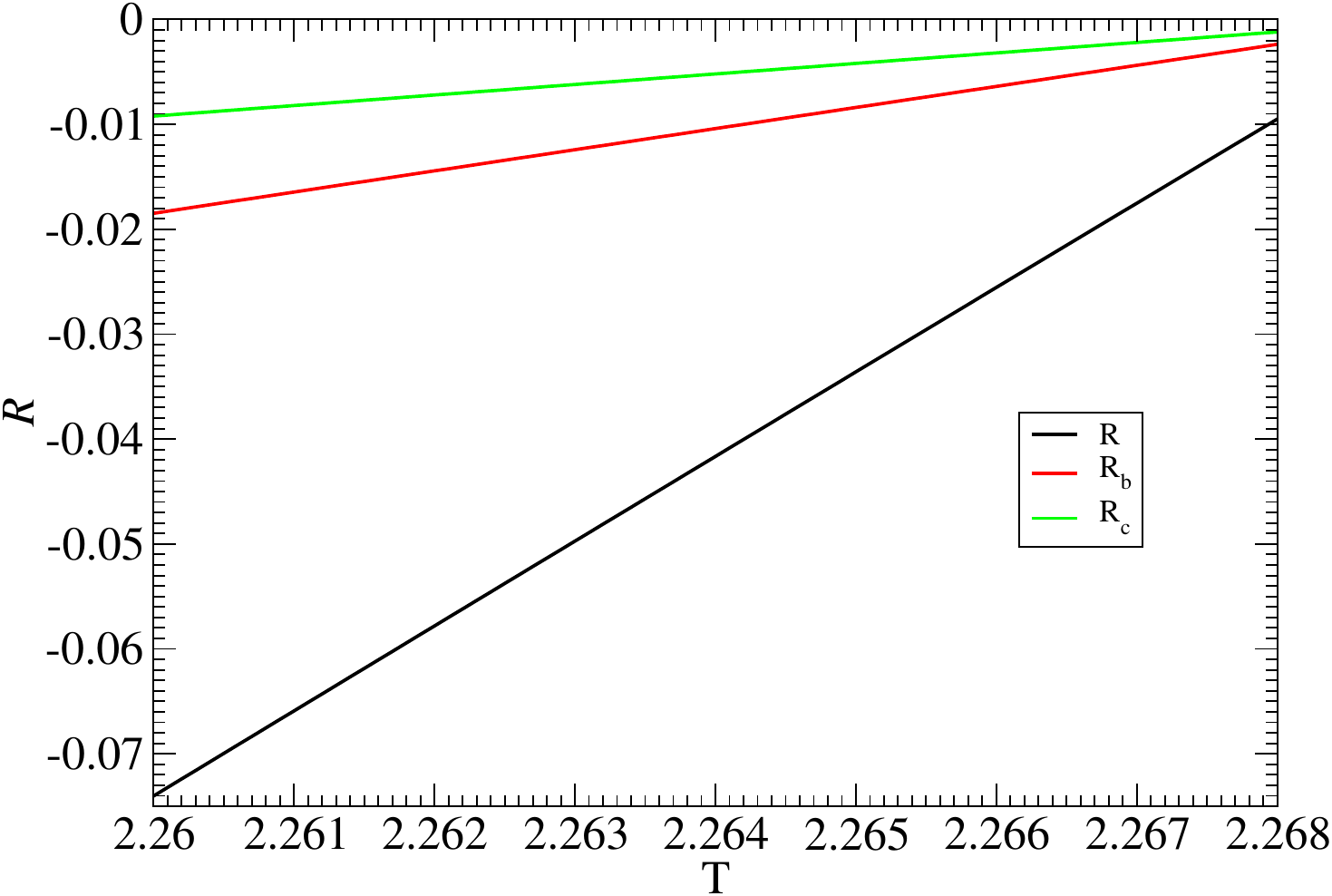}
\caption{Ratio of the magnetization and its first derivative as a function of the temperature. Bulk magnetization $R$ (black curve), border magnetization $R_b$ (red curve), corner magnetization $R_c$ (green curve). Calculated for $m = 300$.}
\label{fig:pomer}
\end{center}
\end{figure}

To achieve this, we first modify the Corner Transfer Matrix Renormalization Group (CTMRG) algorithm to compute magnetization at system boundaries and corners, as well as its first derivative. Second, we adapt the Boundary TRG (BTRG) method~\cite{iino2019} to specifically analyze corner properties. We validate our approach against known exact solutions for the bulk and boundary, and present new, high-precision results for the corner magnetization and its associated critical exponent.

\begin{figure}[h!]
\begin{center}
\includegraphics[width=0.45\textwidth,clip]{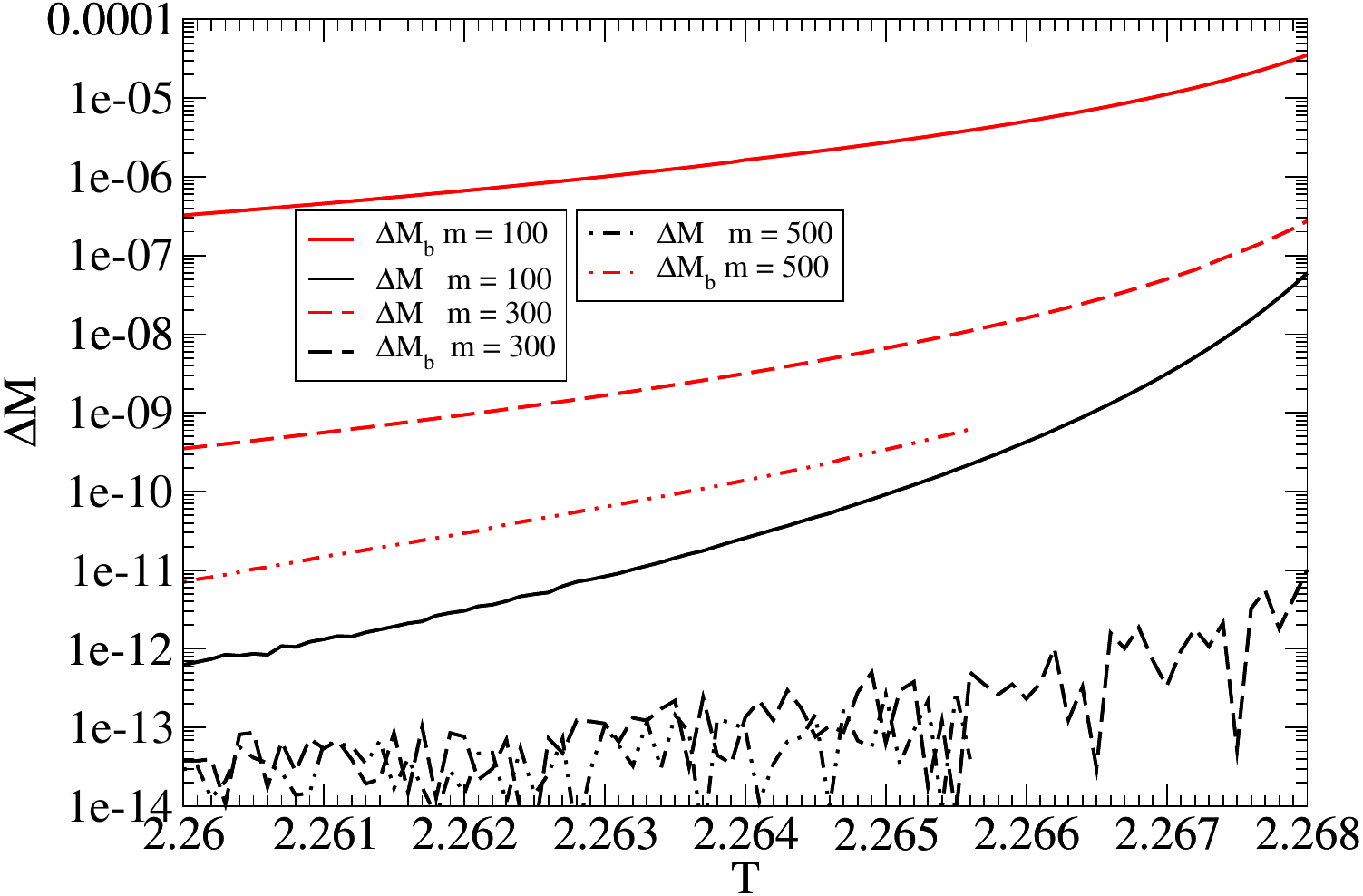}
\caption{Difference between exact and calculated values of magnetization $M$ for bulk (black lines) and boundary (red lines). And for different numbers of states $m = 100$ (full line), $m = 300$ (dashed line), and $m = 500$ (dash-dotted line).}
\label{fig:delM}
\end{center}
\end{figure}

\section{Numerical results}

We are calculating magnetization of Ising model
\begin{equation}
H = -\sum_{\langle i,j\rangle} \sigma_i\sigma_j,
\end{equation}
on the square lattice, and $\sigma_k \in \{0,1\}$ are classical spins. Position indices $i$ and $j$ are shorthand for the two component vectors of the form $i=(i_x,i_y)$ which points form the corner of the lattice $(1,1)$. Magnetization  
\begin{equation}
M_{(i_x,i_y)} = \langle \sigma_{(i_x,i_y)}\rangle,
\end{equation}
is calculated at three positions. In the center of the system $M = \langle\sigma_{N/2,N/2}\rangle$, in the middle of the border $M_b = \langle\sigma_{1,N/2}\rangle$ and in the corner of the system $M_c = \langle\sigma_{1,1}\rangle$.

Calculations are done by modified version of vertex CTMRG ~\cite{vertexctmrg1,vertexctmrg2}. Technical details are discussed in the appendix~\ref{CTMRG}. Magnetization and its first derivative with respect to the temperature are calculated in order to remove the leading non-analytical part of its temperature dependence and for better extraction of the critical exponent $\beta$ and critical temperature $T_C$. In order to extract critical exponents $\beta$ we calculate ratios
\begin{equation}
R = \frac{M}{M^{\prime}}.
\end{equation}

\begin{figure}[h!]
\begin{center}
\includegraphics[width=0.45\textwidth,clip]{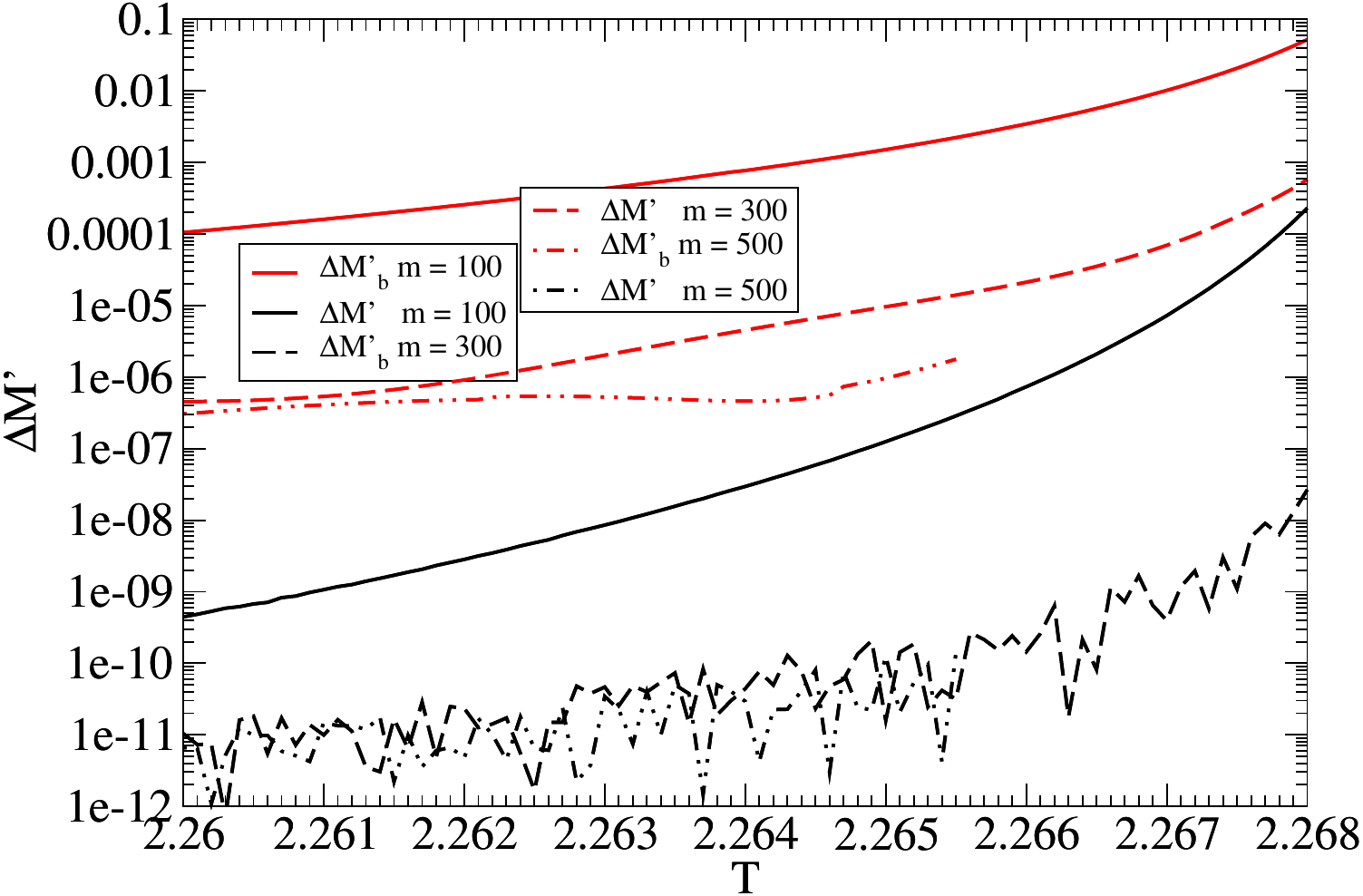}
\caption{Difference between exact and calculated values of the first derivative of magnetization $M$ with respect to temperature $T$ $M^{\prime}$ for bulk (black lines) and boundary (red lines). And for different numbers of states $m = 100$ (full line), $m = 300$ (dashed line), and $m = 500$ (dash-dotted line).}
\label{fig:delMp}
\end{center}
\end{figure}

Critical exponents can be extracted by fitting temperature dependence of the ratios around critical temperature using the fitting function
\begin{equation}\label{rfit}
R(T) =  -\frac1\beta\left(T_C - T\right) + c\left(T_C - T\right)^2,
\end{equation}
where $\beta$, $T_C$ and $c$ are fitting parameters.

Exact formulas for magnetization of the Ising model are known both for central spin $M$ ~\cite{yang} 
\begin{eqnarray}\label{exactbulk}
M &= \left[ 1- \frac1{\sinh^4\left(\frac1T\right)}\right]^{\frac18},\\
M^{\prime} = \frac{dM}{dT} &=  \frac1{T^2M^7}\frac{\cosh\left(\frac2T\right)}{\sinh^5\left(\frac2T\right)},
\end{eqnarray}
and for border spin $M_b$ \cite{mccoy}
\begin{eqnarray}\label{exactcorner}
M_b &= \sqrt{\frac{\cosh\left(\frac2T\right) - \coth\left(\frac2T\right)}{\cosh\left(\frac2T\right)-1}},\\
M_b^{\prime} = \frac{dM_b}{dT} &= \frac{1}{T^2M_b}\frac{\sinh^3\left(\frac2T\right)+1-\cosh^3\left(\frac2T\right)}{\sinh^2\left(\frac2T\right)\left[\cosh\left(\frac2T\right)-1\right]^2}.
\end{eqnarray}

Magnetization is depicted in Fig.~\ref{fig:mag}, its first derivative at Fig.~\ref{fig:der}, and their ratio in Fig.~\ref{fig:pomer}. We decided to calculate them in the interval of temperatures $T\in\langle 2.26,2.268\rangle$ with the temperature step $\Delta T= 0.0001$. It was chosen because it is close enough to critical temperature $T_C = \frac{2}{\ln(1 + \sqrt(2) )} \approx 2.269185314213022$ so that we can rely on the first two terms in the Taylor series of the ratio $R$ and also not too close so that our results are reliable enough. We do not depict the exact values of these variables because, within the range of the graph, they are identical to the calculated points as can be seen from Figs~\ref{fig:delM},\ref{fig:delMp},\ref{fig:delR}. Precision of our method is discused more in appendix~\ref{precision}.

\begin{figure}[ht!]
\begin{center}
\includegraphics[width=0.45\textwidth,clip]{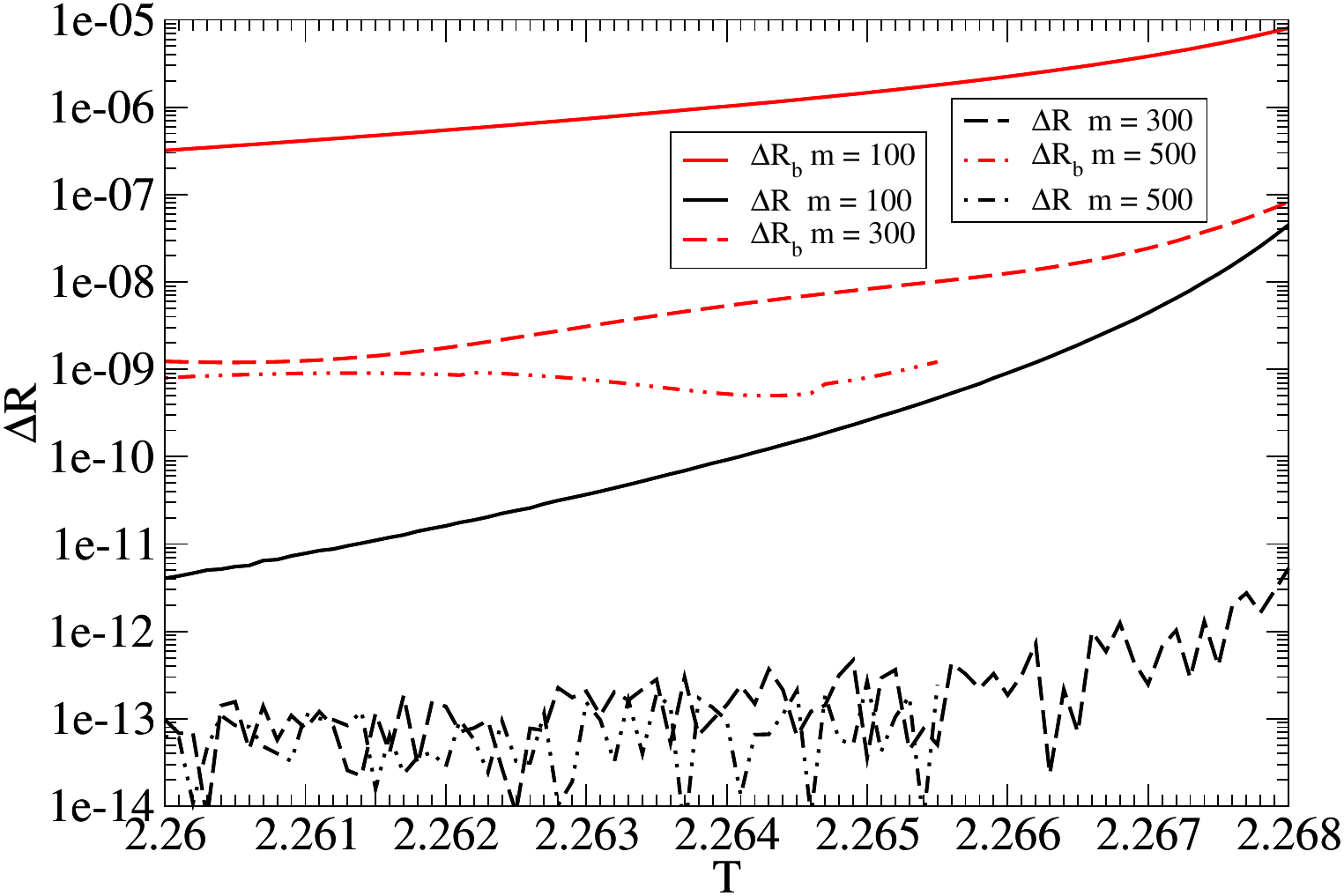}
\caption{Difference between exact and calculated values of the ratio of magnetization $M$ and the first derivative of magnetization $M$ with respect to temperature $T$ $M^{\prime}$ $R$ for bulk (black lines) and boundary (red lines), where $m = 100$ (full line), $m = 300$ (dashed line) and $m = 500$ (dash-dotted line).}
\label{fig:delR}
\end{center}
\end{figure}

 Using fitting function~\ref{rfit} for precision $m = 1000$ and for a range of temperatures $T\in\left\langle2.26,2.268\right\rangle$ we will obtain values for $\beta$ and $T_C$. Ratios are depicted in Fig.~\ref{fig:pomer}.  Fitting values for critical exponents $\beta$ and critical temperature $T_C$ are collected in Tab.~\ref{tabulka1}. Errors shown there are fitting errors. Fitting values of the coefficient $c$ are $c = -6.45(1)$, $c_b = -1.285(2)$ and $c_c = -0.375(7)$ We can see that deviation from known results is at the fifth decimal place or better. We expect that results for the ratio for corners will be slightly less precise.

\begin{table}[htb]
\begin{tabular}{| c | l | l |}
\hline
position & $\beta$ & $T_C$ \\
\hline\hline
center      & 0.125007(2) & 2.26918539(5) \\
\hline
boundary & 0.500020(7) & 2.26918537(3) \\
\hline
corner      & 1.00018(8) & 2.2691859(2) \\
\hline
\end{tabular}
\caption{Fitting parameters of the ratio $R$  for different positions on the lattice for the precision $ m = 1000$. Known exact values are $\beta = 0.125$, $\beta_b = 0.5$ and $T_C\approx 2.269185314213022$.}
\label{tabulka1}
\end{table}

\section{Conclusion}

We have shown that our method is precise enough for high-precision determination of the critical exponent $\beta$. Results for the known values of $\beta$ for bulk and at the border of the system differ from the exact values at the fifth or sixth decimal place . The method strongly suggests that critical exponent $\beta_c = 1$ in the corner of the classical two-dimensional Ising model.

\begin{acknowledgments}
The support received from the project PRESPEED APVV-20-0150 and VEGA Grant QUASIMODO
 VEGA 2/0156/22 is acknowledged. 
 Funded by the EU NextGenerationEU through the Recovery and Resilience Plan for Slovakia under the project No. 09I03-03-V04-00682.
\end{acknowledgments}

\appendix

\section{Method CTMRG}\label{CTMRG}

\begin{figure}[t!]
\begin{center}
\includegraphics[width=0.45\textwidth,clip]{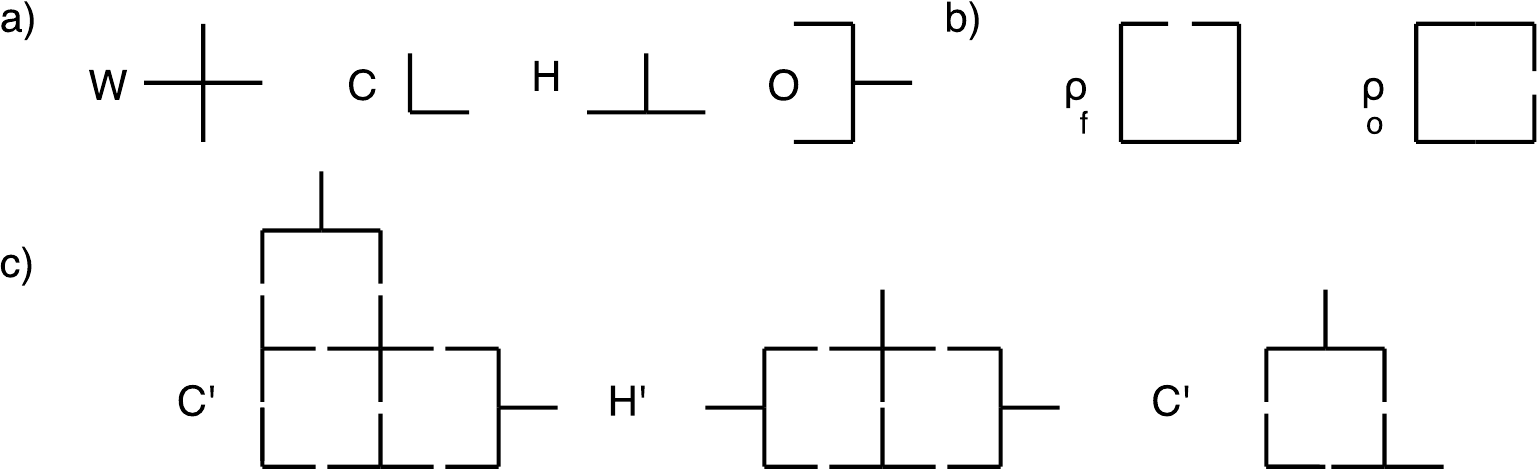}
\caption{Graphical representation of modified CTMRG.In part a basic tensors are presented. In part b schematic representation of density matrices $\rho_f$ and $\rho_o$ are shown. In part c renormalization process for new iterations for corner transfer matrix $C$, half transfer matrix $H$ and boundary corner transfer matrix $C_b$ is schematically depicted.  }
\label{fig:CTMRG}
\end{center}
\end{figure}

Traditional symmetric CTMRG uses three tensors, Boltzmann weight $W$, half transfer matrix $H$, and corner transfer matrix $C$. In addition to these tensors during the process of renormalization one more tensor arises, orthogonal projector $O$, which consists of eigenvectors of the density matrix. It is used at each iteration step of the method in order to project out least relevant degrees of freedom. 

Traditional procedure is extremely precise for the magnetization in the bulk, but it loses its precision for magnetization at the edge of the system. We presume that the reason is that traditional method enhance the information about the bulk degrees of freedom of the system and suppress the information about boundary degrees of freedom. 

Traditional method expand the system from the center equally in all directions. In order to preserve information about the boundary we decided to grow the system from the bottom edge in a half plane fashion instead of full plane.

In our case the family of tensors consists of one Boltzmann weight $W$, three corner matrices $C_f$, $C_o$, $C_b$, three half transfer matrices $H_f$, $H_o$, $H_0$. This choice of the tensors is dictated by the boundary conditions of our system. We have chosen a mixed boundary conditions. Magnetic field is switched on at the upper edge of the system and rest of the edges are kept field free. Because of this setup corner transfer matrix $C_f$ which has open boundary conditions on one edge and magnetic field on the other is needed and $C_o$ is a symmetric corner transfer matrix that has both edges with open boundary conditions. Similarly, for the half transfer matrices, we have a matrix $H_f$ with the edge with a magnetic field and $H_o$ with the edge with open boundary conditions. In the case of normal renormalization process, the boundary degrees of freedom are suppressed  and the bulk degrees of freedom are enhanced. In order to fight this tendency and salvage the boundary degrees of freedom we have introduced last two tensors. Half transfer matrix tensor $H_0$ which is the unadulterated initial half-row tensor with open boundary conditions and corner transfer matrix $C_b$ which is a non-symmetric matrix with open boundary conditions on both edges, but only one leg is renormalized the other one is kept the same as it was at the beginning. Using these two tensors we are calculating boundary magnetization. Unlike in standard CTMRG during the renormalization process, we create two types of density matrices $\rho_f$ and $\rho_o$. These matrices are schematicaly depicted at Fig.~\ref{fig:CTMRG} part b. From these matrices, we obtain two projectors $O_f$ and $O_o$. Projector $O_f$ is used for the legs with magnetic field, projector $O_o$ for the ones without magnetic field. 

The renormalization process is schematically depicted on Fig.~\ref{fig:CTMRG} part c. Here we show how to construct tensors which belongs to the iteration $i+1$ from the tensors which belongs to the previous iteration $i$. For corner transfer matrix with one leg with the magnetic field $C_f$ the process is depicted on first diagram in part c.  Exact tensors looks like
\begin{equation}
C_{f;i+1} = O_{f;i}H_{f;i}C_{f;i}WH_{o,i}O_{o,i}.    
\end{equation}
Symmetric corner transfer matrix $C_o$ has the same diagram with different tensors
\begin{equation}
    C_{o;i+1} = O_{o;i}H_{o;i}C_{o;i}WH_{o,i}O_{o,i}.
\end{equation}
Similarly for half transfer matrices $H_o$ and $H_f$ we have the same overall process, depicted at the second diagram in part c, but with slightly different tensors. For matrix $H_f$ we have
\begin{equation}
H_{f;i+1} = O_{f;i}H_{f;i}WO_{f;i+1},    
\end{equation}
and for $H_o$
\begin{equation}
H_{o;i+1} = O_{o;i}H_{o;i}WO_{o;i+1}.    
\end{equation}
For the boundary corner transfer matrix $Cb$ process is depicted at last diagram of part c. Exact tensor are
\begin{equation}
C_{b;i+1} = O_{o;i}H_{f;i}C_{b;i}WH_{0}.    
\end{equation}

\section{Method BTR}\label{BTR}
\begin{figure}[htb]
\begin{center}
\includegraphics[width=0.45\textwidth,clip]{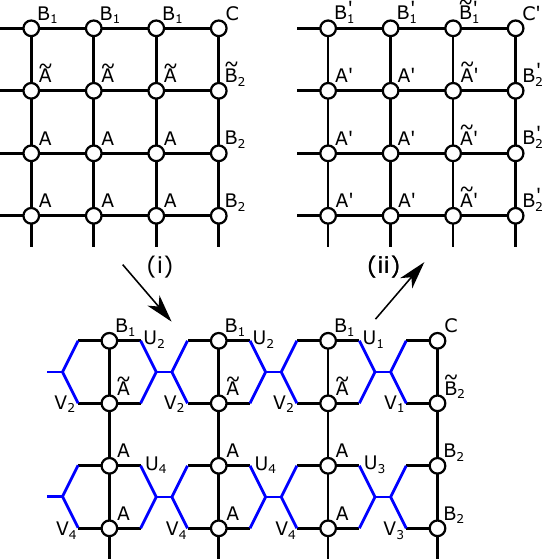}
\caption{
A BTRG renormalization of the tensor-network state along the $y$ axis at the upper-right corner of the square lattice.
(i) The projectors (blue) are inserted into every two horizontal bonds to contract two neighboring tensors.
(ii) Resulting lattice after the vertical contraction. 
}
\label{fig:horisontal}
\end{center}
\end{figure}

\begin{figure}[htb]
\begin{center}
\includegraphics[width=0.45\textwidth,clip]{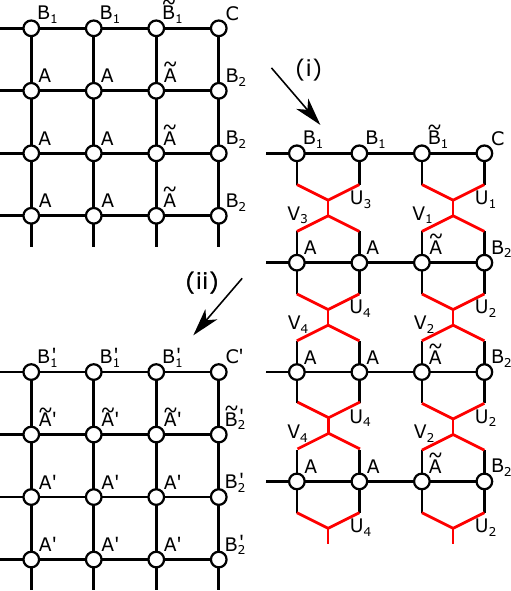}
\caption{
A BTRG renormalization of the tensor-network state along the $x$ axis at the upper-right corner of the square lattice.
(i) The projectors (red) are inserted into every two vertical bonds to contract two neighboring tensors.
(ii) Resulting lattice after the horizontal contraction. 
}
\label{fig:vertical}
\end{center}
\end{figure}

We propose a method to extend and renormalize the upper-right corner of the square lattice.
For this purpose, we modify the BTRG method from Ref.~\cite{iino2019} to include a corner, upper boundary, right boundary, and bulk.
The algorithm performs the renormalization steps alternatively along the $y$ axis and $x$ axis as depicted in Fig.~\ref{fig:horisontal} and Fig.~\ref{fig:vertical}, respectively. 
Before the renormalization step along the $y$ axis (or after the renormalization step along the $x$ axis), the system is represented by one rank-2 corner ($C$), three rank-3 boundary tensors ($B_1$, $B_2$, and intermediate $\tilde{B_2}$), one rank-4 bulk tensor ($A$), and, finally, one rank-4 intermediate tensor ($\tilde{A}$), see Fig.~\ref{fig:horisontal}. 
In step (i) in Fig.~\ref{fig:horisontal}, the BTRG-style projectors $U$ and $V$ are inserted into every two horizontal bonds to contract two neighboring tensors.
In step (ii), the projectors are absorbed into the vertical pairs of tensors.
For concreteness, a new corner tensor $C'$ is obtained as $C' = V_1 (C\tilde{B_2})$. 
Notice, that the intermediate tensors (denoted by a tilde) are now arranged into a second-to-last column whereas before the intermediate tensors were arranged into a row (second from up). 
For example, now we have two intermediate tensors $\tilde{{B'}_{1}} = V_2 (B_1 \tilde{A}) U_1$ and $\tilde{A'} = V_4 (A A) U_3$.
The new boundary tensors are ${B'}_1 = V_2 (B_1 \tilde{A}) U_2$ and ${B'}_2 = V_3 (B_2 B_2)$.
Lastly, the new bulk tensor $A'$ is defined as $A' = V_4 (A A) U_4$. 
The renormalization step along the $x$ axis is performed analogously and it starts from the tensor arrangement which resulted from the renormalization step along the $y$ axis, see Fig.~\ref{fig:vertical}. 
In step (i) in Fig.~\ref{fig:vertical}, the BTRG-style projectors $U$ and $V$ are inserted into every two vertical bonds to contract two neighboring tensors.
In step (ii), the projectors are absorbed into the horizontal pairs of tensors.
For concreteness, a new corner tensor $C'$ is obtained as $C' = U_1 (\tilde{B_1} C)$. 
We also obtain new intermediate tensors $\tilde{A'}$ and $\tilde{{B'}}_2$ as $\tilde{A'} = V_3 (A A) U_4$ and $\tilde{B'}_2 = V_1 (\tilde{A} B_2) U_2$.
The new boundary tensors are defined as ${B'}_1 = U_3 (B_1 B_1)$ and ${B'}_2 = V_2 (\tilde{A} B_2) U_2$. 
Lastly, the new bulk tensor $A'$ is defined as $A' = V_4 (A A) U_4$. 

\section{Precision of the method}\label{precision}
In order to test precision of our method, magnetization its first derivative and their ratio are compared to either to exact formulas~(\ref{exactbulk}-\ref{exactcorner}) or we compare our two methods to each other~\ref{CTMRG} \ref{BTR}. 
Error of the ratio as a function of the errors in the magnetization and its first derivative is bounded from above

\begin{widetext}
\begin{eqnarray}
\Delta R = \sqrt{\left(\frac{\Delta M}{M^\prime}\right)^2 + \left(\frac{M\Delta M^\prime}{M^{\prime2}}\right)^2+2r\frac{M\Delta M\Delta M^\prime}{M^{\prime3}} } \le \nonumber\\
\frac1{\left| M^\prime\right|}\left(\left| \Delta M\right| + \left|\frac{M}{M^{\prime}}\right|\left|\Delta M^{\prime}\right|\right)
\end{eqnarray}
\end{widetext}

where $\Delta R$ is error of the ratio $R$, $\Delta M$ is error of the magnetization $M$, $\Delta M^\prime$ is error of its first derivative $M^\prime$ and $r$ is correlation coefficient between $M$ and $M^\prime$. For some reason, our automatic differentiation produces a larger error for the derivative of the magnetization when compared to the magnetization. The difference is around two orders of magnitude.

As we can see form fig.~\ref{fig:der} first derivative of magnetization is relatively large for bulk and boundary case. For them for the smaller temperatures major source of error comes from the error of the derivative but closer to the critical point prefactor of the derivative error will decrease and the dominant source of error will become the error in the magnetization. In the corner case we expect that both contributions will be comparable.

The precision of our method can be assessed by comparing our results with exact values. Difference is shown on Figs.~\ref{fig:delM},~\ref{fig:delMp},~\ref{fig:delR}. In Fig.~\ref{fig:delM} difference between exact and calculated magnetization for various $m = 100,300,500$. As expected for $m=300$ and $m=500$, the precision of the bulk results is very close to our numerical convergence criterion which means we will stop the calculation when two consecutive results are closer than $10^{-14}$. In Fig.~\ref{fig:delMp} difference between the exact and calculated first derivative of magnetization with respect to temperature is depicted. Errors are approximately two orders of magnitude bigger, but the overall picture is the same as for magnetization errors. Errors of the ratio $R$ are depicted in Fig.~\ref{fig:delR} these results are comparable to results for magnetization.

Errors of the ration $R_b$ between boundary magnetization $M_b$ and its first derivative $M^{\prime}_b$ are close to $10^{-9}$ for whole interval of temperatures. Therefore we can be confident that our results are very accurate as can be seen from our estimations of $\beta$ and $T_C$ see Tab.~\ref{tabulka1}. 

We are not sure if this accuracy will translate into our results for the calculations at the corner. In order to asses precision of these calculations we have compared our methods and also studied convergence of our results with increasing number of states kept $m$. 

\bibliography{rohovectmrg}

\end{document}